# Trans-scale spin Seebeck effect in nanostructured bulk composites based on magnetic insulator

Sang J. Park[1,*], Keisuke Hirata[2,3], Hossein Sepehri-Amin[1], Fuyuki Ando[1], Takamasa Hirai[1], and Ken-ichi Uchida[1,3,*]

[1] National Institute for Materials Science, Tsukuba 305-0047, Japan

[2] Toyota Technological Institute, Nagoya 468-8511, Japan

[3] Department of Advanced Materials Science, Graduate School of Frontier Sciences, The University of Tokyo, Kashiwa 277-8561, Japan

*Correspondence to: PARK.SangJun@nims.go.jp (S.J.P.); UCHIDA.Kenichi@nims.go.jp (K.U.)

## Abstract

The spin Seebeck effect enables thermoelectric conversion through thermally generated spin currents in magnetic materials, offering a promising transverse geometry for scalable devices. However, conventional spin Seebeck devices are confined to nanoscale thin-film architectures, with significantly restricted output power due to the intrinsic constraints of spin and magnon diffusion lengths. Here, we demonstrate a trans-scale spin Seebeck effect using nanostructured bulk composites composed of Pt-coated yttrium iron garnet powders fabricated via dynamic powder sputtering and low-temperature sintering. The resulting three-dimensional composites exhibit continuous Pt channels and robust mechanical integrity. Transverse thermoelectric measurements reveal isotropic spin Seebeck signals at the bulk scale. Power analysis indicates that the three-dimensional architecture enables scalable volumetric thermoelectric power generation beyond diffusion-limited thin-film spin Seebeck geometries. This work establishes a scalable platform for spin Seebeck thermoelectric conversion, bridging nanoscale spin caloritronics with macroscopic device integration.

## Introduction

The spin Seebeck effect (SSE) enables the generation of electrical voltage via thermally driven spin currents (or magnons) in magnetically ordered materials[1–4]. In the SSE, a temperature gradient applied across a ferro(i)magnetic material (FM) excites spin currents, which are then pumped into an adjacent normal metal (NM). There, they are converted into a transverse electric field through the inverse spin Hall effect (ISHE)[5,6]. Owing to the orthogonal relationship between the heat flux and the induced electric field, the SSE is classified as a transverse thermoelectric effect[7–10]. Since its discovery in 2008[11], the SSE has been a central topic of interest in both spintronics and thermoelectrics, serving as a platform for investigating fundamental transport physics[1–3,8,9,11–20] and as a basis for thermoelectric energy harvesting applications[1–3,7–9,11,21–27], based on thermally generated spin currents.

SSE-based thermoelectric energy harvesting has been proposed as a solution to the limitations of conventional thermoelectric devices, which rely on the longitudinal Seebeck effect to generate a charge current along the direction of a temperature gradient[28–31]. These longitudinal devices typically adopt π-shaped geometries with multiple n-type and p-type thermoelectric legs interconnected by electrodes such as solder. Such complex architectures hinder manufacturability, scalability, and overall device-level efficiency[7,32]. In contrast, SSE devices utilize simplified transverse configurations that eliminate the need for electrically and thermally sensitive junctions (Fig. 1a)[7–9]. This architectural simplicity enhances the mechanical robustness and scalability of devices, while also enabling more flexible material design. More importantly, the SSE enables thermoelectric conversion even in ferro(i)magnetic insulators (FMI), which are inaccessible to conventional thermoelectrics relying on mobile charge carriers. This unique capability allows SSE-based devices to harvest waste heat from insulating materials, thereby expanding the range of usable thermal sources beyond that of traditional approaches. In particular, the independent generation and propagation of spin and charge currents in the FM and NM layers, respectively, allow for separate optimization of each material, circumventing intrinsic trade-offs in material parameters, such as those dictated by the Wiedemann-Franz law[28]. These advantages make SSE-based devices promising candidates for scalable and reliable thermoelectric energy harvesting, despite their relatively low thermopower to date.

Over the past decades, numerous efforts have been made to enhance the SSE by improving energy conversion efficiencies within the FM and NM, and at their interface, for example, through magnon temperature manipulation[16], the use of topological materials[33], and interfacial engineering[22,34]. However, the practical implementation of SSE remains challenging, primarily due to the limited power output resulting from the hetero-structured nature of thin-film FM/NM devices (Fig. 1a). The device thickness is fundamentally constrained by the characteristic length of spin transport, namely the spin diffusion length ($\lambda_s$), over which the spin angular momentum decays. As a result, the optimal thickness of the NM layer is typically limited to a few nanometers (e.g., less than 5 nm for sputtered Pt[35]), leading to high internal resistance and reduced output power.

The effective FM thickness is similarly governed by the magnon diffusion length ($\lambda_m$) particularly in FMI, where spin angular momentum is carried solely by magnons, i.e., collective dynamics of localized magnetic moments[36–38]. For instance, yttrium iron garnet (YIG), a widely studied FMI known for its exceptionally low Gilbert damping constant ($<10^{-4}$)[39], exhibits $\lambda_m$ of ~10 μm at room temperature[40,41], which is significantly longer than that of ferro(i)magnetic metals (FMMs), e.g., 3.5 nm for sputtered Ni[42]. Nevertheless, these length scales remain microscopic, limiting device configurations to quasi-two-dimensional thin-film geometries (Fig. 1a), where only narrow volumes of FM and NM adjacent to the interface contribute to the SSE (see Fig. 1c). In addition, this planar architecture introduces operational anisotropy; the heat flux and magnetization of FM must be in the directions perpendicular and parallel to the FM/NM interface, respectively, thereby restricting practical applicability. To overcome these structural limitations and enable scalable SSE-based energy harvesting, a transition from thin-film architectures to bulk-compatible, three-dimensional geometries is essential, which we refer to as trans-scaling.

One approach to implementing trans-scaling is to engineer extrinsic device parameters. Specifically, a three-dimensional nanostructured composite material[24,43] can be constructed by embedding multiple NM-coated FM domains throughout the bulk (Fig. 1b). This design effectively extends conventional quasi-two-dimensional SSE devices into bulk composites, increasing the usable thermal volume, which enhances the overall heat-to-electricity conversion capability without relying on high-resistance thin-film architectures. This concept, referred to as the trans-scale SSE, leverages composite architectures in which NM channels are distributed throughout the FM matrix, allowing volumetric utilization of spin currents and circumventing the intrinsic limitations imposed by $\lambda_s$ and $\lambda_m$ (Fig. 1d).

To elucidate the physical basis that enables the trans-scale SSE to survive such a geometrically complex FM/NM interface network, we first recall that in an FM/NM bilayer the ISHE field obeys

$$\mathbf{E}_{\mathrm{ISHE}} \propto \mathbf{j}_s \times \hat{m} \tag{1}$$

where $\mathbf{j}_s$ is the spatial direction of the thermally driven spin current induced by a temperature gradient $\nabla T$, and $\hat{m}$ is the magnetization (spin-polarization) direction. For an arbitrarily oriented interface patch with unit normal $\hat{n}$, the injected spin current flows along $\hat{n}$, and its magnitude scales with the normal projection of $\nabla T$. The local ISHE response therefore generalizes to

$$\mathbf{E}_{\mathrm{ISHE}}(\hat{n}) \propto (\hat{n} \cdot \nabla T)(\hat{n} \times \hat{m}). \quad (2)$$

Because both prefactors in Equation (2) reverse sign under inversion of the interface normal ($\hat{n} \rightarrow -\hat{n}$), their product remains invariant, ensuring that rough, concave, convex, or oppositely oriented facets contribute constructively rather than cancelling. When the orientations of $\hat{n}$ are statistically isotropic in three-dimensional space, the orientational average reduces to

$$\langle (\hat{n} \cdot \nabla T)(\hat{n} \times \hat{m}) \rangle = \frac{1}{3}(\nabla T \times \hat{m}) \quad (3)$$

indicating that the emergent macroscopic SSE retains the usual ISHE symmetry $\nabla T \times \hat{m}$ independent of morphological complexity, with a geometric reduction factor of 1/3. A full derivation is provided in Supplementary Note 1.

Despite its conceptual promise, the realization and development of such three-dimensional SSE devices remain challenging due to difficulties in NM-FM coating and sintering. First, conventional methods for coating FM with NM, such as atomic layer deposition (ALD), rely on chemical precursors and often produce poor FM/NM interfacial quality due to residual organic contaminants and weak mechanical FM/NM bonding. These interfacial problems significantly degrade the spin mixing conductance; for example, YIG/Pt interfaces prepared by ALD exhibit spin mixing conductance reduced by a factor of ~20 compared to those prepared by sputtering[44]. FMMs, while easier to fabricate into NM microchannel structures, exhibit very short $\lambda_s$ on the order of a few nanometers[42] and are susceptible to oxidation, which limits their effectiveness in such composites[43]. Moreover, electrical shunting between the FMM and NM can substantially reduce the SSE signal generated in the NM layer. The anomalous Nernst effect (ANE)[10,45] generated in the FMM can also interfere with the detection of the pure SSE contribution. Therefore, oxide-based FMIs with long $\lambda_m$, such as YIG, are more desirable. However, they are structurally more difficult to process due to their thermodynamic stability, which requires high temperatures for sintering (e.g., above 900 °C for YIG[46]). Such high temperatures can lead to several issues in the thin NM layer[47,48], including increased surface roughness, film discontinuity, and dewetting, all of which degrade layer quality (see Supplementary Note 2 for detailed discussion). Overcoming these challenges is thus crucial for realizing SSE-based energy harvesting devices that can effectively utilize heat from insulators at macroscopic scales. While the conceptual structure of bulk SSE composites was outlined nearly a decade ago[43], their practical realization has remained difficult due to the intrinsic constraints discussed above. The successful demonstration of such systems would represent a significant advance toward practical spin-caloritronic applications.

## Results

### Fabrication of YIG-Pt composite materials

Here, we demonstrate the trans-scale SSE using FMI-NM composite materials composed of multiple YIG-Pt domains distributed throughout the bulk (Fig. 1b). To fabricate this structure, we established a dynamic powder sputtering system[49] (Supplementary Fig. 1 for photographic setup) that enables uniform, nanometer-scale Pt coating on YIG powders without the use of chemical precursors (Fig. 2 and Methods). Uniform Pt coating was achieved by simultaneously rotating and vibrating the YIG powders during deposition. Figure 2b shows a transmission electron microscopy (TEM) image of Pt-coated YIG powders, confirming the formation of a continuous 5-nm-thick Pt layer. When the Pt thickness was reduced below 5 nm, the coating became non-uniform (Supplementary Fig. 2).

This metallic coating not only provides spin-caloritronic functionality (i.e., thermal spin pumping), but also improves mechanical adhesion between powders owing to the ductility and high atomic mobility of metals compared to oxides. As a result, the Pt-coated powders could be sintered through metal-mediated ductile channels at low temperatures, including 300 °C and even at room temperature (Methods), while maintaining the conducting paths in the bulk composite. In contrast, pure YIG powders could not be densified under the same conditions, highlighting the critical role of Pt layers in enabling densification.

Scanning electron microscope (SEM) images reveal the formation of micrometer-scale Pt channels, which increase the effective volume of both FM and NM materials contributing to the SSE (Fig. 2c). X-ray diffraction (XRD) confirmed that the YIG remains the predominant phase without the formation of secondary phases during fabrication (see Supplementary Note 3 and Supplementary Fig. 3 for detailed discussion). A more comprehensive discussion of potential contributions from secondary phases is presented in Discussion, where we also examine the exclusion of possible signal contamination from parasitic thermoelectric effects.

Four composite samples were prepared to experimentally demonstrate the trans-scale SSE. These samples differed in a Pt thickness (15 nm and 30 nm), both of which are well above $\lambda_s$ of Pt (<5 nm[35]), ensuring clear observation of the SSE without discontinuities in the Pt channels. In addition, two different sintering conditions were employed: pressing at 300 °C and room temperature under high pressure (**Methods**). The sintered pellets were labeled according to their Pt thickness and sintering temperature (i.e., 15-300, 15-RT, 30-300, and 30-RT), and they exhibited well-consolidated structures with relative densities ranging from 67% (15-RT) to 73–75% (15-300, 30-300, and 30-RT), as shown in Supplementary Fig. 4.

**Electrical and thermal conductivity of nanostructured bulk composites**

We then evaluated the electrical conductivity ($\sigma$) of the bulk composite by normalizing the volumetric electrical conductance with the geometric dimensions of the composite, including YIG and Pt (Methods). The measured $\sigma$ values range from (5–8) × $10^2$ S/m for 15 nm Pt samples and (23–50) × $10^2$ S/m for 30 nm Pt samples at room temperature (Supplementary Fig. 4a), all of which fall within the "bad metal" regime[50]. These values are higher than those of typical homogeneous semiconductors (non-doped Si ~ 1.5 × $10^{-3}$ S/m and Ge ~ 2 S/m) and oxide-based bulk ANE materials ((3–26) × $10^2$ S/m)[51–53].

Next, we estimated the thermal conductivity ($\kappa$) by multiplying the measured thermal diffusivity, specific heat, and density (Supplementary Fig. 4). The $\kappa$ values are 1.46–1.87 W/mK, which are substantially lower than those reported for polycrystalline YIG (~4.9 W/mK) and single-crystalline YIG (~7–8 W/mK)[46,54]. Such low $\kappa$, corresponding to 30–38 % of the polycrystalline reference, cannot be attributed solely to the relative density, for example, 75% for 30-RT. The reduction likely reflects multiple phonon-scattering channels introduced by the composite structure, including porosity, grain boundaries, interface defects, and thermal boundary resistance at the YIG/Pt interfaces. In insulating YIG, heat conduction is mainly governed by phonons, with a relatively small magnon contribution expected at room temperature in polycrystalline samples[46,54]. Therefore, controlling the YIG/Pt interfacial density, defect structure, and pore morphology may provide an additional route for thermal-conductivity engineering in bulk SSE composites.

**Observation of trans-scale SSE in nanostructured bulk composites**

We now present the observation of the trans-scale SSE in the YIG-Pt composites under two distinct magnetothermal configurations: dubbed "$H_z$-$\nabla_x T$" ("$H_x$-$\nabla_z T$") configuration measuring voltage into the $y$-direction, while the orthogonal $H$ is applied to $z$ ($x$)-direction and temperature gradient ($\nabla T$) is applied to $x$ ($z$)-direction, as shown in Fig. 3a (3d). While rotating the magnetothermal configurations are not expected to generate a consistent SSE signal in conventional FM/NM planar film structures[14,20] due to lack of spin pumping under out-of-plane $H$, we show that both geometries consistently produce SSE signals in our composites, highlighting a distinctive feature of isotropic trans-scale SSE. In the following, we provide a systematic evaluation to rigorously validate the observation by excluding potential parasitic contributions that could influence voltage measurement and by assessing the effects of secondary phases. These parasitic signals include the Seebeck effect driven by oblique heat flux[55,56], the (proximity-induced) ANE[14,20,57], and magnetoresistance effects.

We first focus on the observed transverse electric field ($E_y$) in the $H_z$-$\nabla_x T$ configuration (Fig. 3a). The signal was symmetrized with respect to $H$ to remove field-even components, such as the magneto-Seebeck effect caused by unintended transverse heat gradients ($\nabla_y T$) in Pt. As shown in Fig. 3b, a clear field-odd signal was observed in the 30-RT sample within the $H_z$ range of $\pm 5$ kOe. The linear dependence of $E_y$ on $\nabla_x T$ was confirmed in Fig. 3c, using $E_y^{\mathrm{sat}}$ denoting $E_y$ at the saturated fields (>2.5 kOe). These results directly confirm that the observed signal originates from magnetism-induced transverse thermoelectric conversion, with longitudinal thermoelectric and magneto-electric contributions effectively excluded. The linear slope of Fig. 3c corresponds to the spin Seebeck coefficient ($S_{SSE} \equiv E_y^{sat}/\nabla_x T$), shown in Supplementary Fig. 5. Samples sintered at room temperature exhibited $S_{SSE}$ values of 17.3 nV/K (15-RT) and 14.8 nV/K (30-RT), which are higher than those of samples sintered at 300 ℃ (9.2 nV/K for 15-300 and 7.5 nV/K for 30-300). The 15-nm-thick Pt samples exhibited slightly higher $S_{SSE}$ than the 30-nm-thick Pt samples, qualitatively consistent with the typical Pt thickness dependence of $S_{SSE}$ in the conventional YIG/Pt systems due to the small $\lambda_s$ of Pt[35]. Using the estimated $\sigma$ and $S_{SSE}$, we calculated the transverse power factor ($PF_{SSE} = \sigma S_{SSE}^2$) driven by the SSE. Supplementary Figure 5b shows that 30-RT exhibited the largest $PF_{SSE}$ (5.1 × $10^{-13}$ W/mK$^2$), approximately 630% higher than that of 15-300 (0.7 × $10^{-13}$ W/mK$^2$), strongly suggesting the importance of optimizing the Pt thickness and sintering conditions to enhance the transverse thermoelectric performance.

Next, we demonstrate the SSE in the $H_x$-$\nabla_z T$ configuration using the 30-RT sample by rotating it 90 degrees to the original magnetothermal configuration, where the $E_y$ signal was measured under $H_x$ and $\nabla_z T$ (Fig. 3d). In conventional anisotropic FM/NM thin-film structures, no SSE signal is expected in this rotated magnetothermal configuration, due to the absence of spin pumping at the FM/NM interface. Only (proximity-induced) ANE can be captured in such cases[14,20]. Figure 3e shows a clear field-odd signal, with $S_{SSE}$ measured as 14.4 nV/K, which is consistent with that obtained in the $H_z$-$\nabla_x T$ configuration (Fig. 3c and Supplementary Fig. 5). This consistency supports the assignment of the signal to SSE and demonstrates the reproducibility of the bulk SSE response under orthogonal magnetothermal configurations, while confirming the directional isotropy of the system and its potential for energy harvesting applications.

### Validation of trans-scale SSE and exclusion of parasitic effects

One may raise concerns about parasitic contributions from the (proximity-induced) ANE, which can arise in both $H_z$-$\nabla_x T$ and $H_x$-$\nabla_z T$ configurations, if secondary (proximity-induced) FMM phases were unintentionally formed during fabrication. However, we rule this out based on multiple lines of evidence. First, XRD measurements show no detectable formation of FMMs, such as $Fe_3O_4$ or FePt (Supplementary Fig. 3b and Supplementary Note 3), thereby excluding their contributions to ANE. This is further supported by the magnetization data: if such FM phases were present, the saturation magnetization would increase, considering that the ferrimagnetic YIG (37.9 emu/g) exhibits much lower values that those of common Fe-based FMs (85-92 emu/g for $Fe_3O_4$[58], 74-76 emu/g for $\gamma$-$Fe_2O_3$[59], and 186-218 emu/g for $\alpha$-Fe[60]). In contrast, we observe a decrease in saturation magnetization upon the deposition of paramagnetic Pt (Supplementary Fig. 6), and the magnetization curve shows a single, well-defined hysteresis loop without anomalies, confirming the absence of secondary magnetic phases (Supplementary Fig. 6 for detailed low-field data).

We also exclude the contribution of proximity-induced ANE because the magnetic proximity effect between YIG and Pt and its contribution to transverse thermoelectric effect are well known to be negligibly small, as demonstrated by X-ray magnetic circular dichroism and transport studies[14,61–64]. Consistently, SQUID-VSM measurements show no additional magnetic component or anomalous magnetic behavior. Even if local strain-induced interfacial modification exists below the detection limit, it would affect interfacial spin transparency or SSE-driven ISHE efficiency[65] rather than generate an independent ANE-like voltage. To further confirm the absence of such contributions from secondary (proximity-induced) FMMs, we performed Hall measurements, measuring $E_y$ under a charge current $I_z$ and $H_x$, to sensitively detect any (proximity-induced) FMMs effectively contributing to the transport (Fig. 3e). The Hall voltage ($V_y^{\mathrm{Hall}}$) shows a linear response to $H_x$, indicating the absence of (proximity-induced) FMMs effectively contributing the transverse signals. Furthermore, the transverse thermoelectric signal exhibits a suppression at high magnetic fields (Fig. 3f), which is a well-established hallmark of the SSE[13,15]. This suppression arises from the field-induced reduction of thermally excited magnons and is not expected for ANE of FMM impurities, whose transverse voltage simply follows the magnetization and saturates at high fields. This high-field damping therefore provides an additional signature that the observed transverse voltage originates from magnon-driven SSE rather than ANE.

Finally, to provide an independent and unambiguous confirmation, we prepared a YIG–W composite, where W has a negative spin Hall angle. Figure 4 shows that the transverse thermoelectric voltage of the YIG–W composite in the $H_z$-$\nabla_x T$ configuration exhibits a clear sign reversal relative to the YIG–Pt composite under otherwise identical conditions, despite the higher noise level associated with the larger electrical resistance of the control sample (Supplementary Note 4). This inversion provides direct evidence that the observed transverse voltage originates from the magnon-driven thermal spin pumping and its ISHE-driven spin-charge conversion in the bulk composite, establishing the trans-scale SSE.

We additionally note that the SSE originates from magnons generated in YIG, not from secondary magnetic phases, such as antiferromagnetic α-$Fe_2O_3$ and $YFeO_3$. While both are reported to exhibit long-range magnon transport[66,67], their formation could not be distinguished in the XRD data shown in Supplementary Fig. 3 and their magnon transport is known to be strongly anisotropic and to require high magnetic fields. These characteristics suggest that they are inert to magnon transport within the field range used in our measurements.

**Thickness-dependent power scaling enabled by volumetric spin conversion**

We evaluate the device-level performance using the maximum output power,

$$P_{\text{max}} = \frac{V_{\text{oc}}^2}{4R_{\text{int}}}, \quad (4)$$

where $V_{\text{oc}}$ is the open-circuit voltage and $R_{\text{int}}$ is the internal resistance of the device. To isolate the architectural contribution, we fix the lateral device dimensions ($L_y$ and $L_z$) and the applied heat-flux boundary condition, while varying only the device or FM thickness $t$ (Fig. 5).

For conventional quasi-two-dimensional interfacial SSE devices (Fig. 1a), the open-circuit voltage follows $V_{\text{oc}}^{2\text{D}}(t) \propto S_0(1 - e^{-t/\lambda_m})$. Under fixed NM geometry, the internal resistance remains approximately thickness independent. Therefore,

$$P_{\text{max}}^{2\text{D}}(t) \propto (1 - e^{-t/\lambda_m})^2, \quad (5)$$

which saturates for $t \gg \lambda_{\text{m}}$. In the present analysis a representative $\lambda_{\text{m}}$ = 10 μm, as reported for YIG at room temperature[17,68], was adopted for the 2D reference calculations.

In contrast, for the present quasi-three-dimensional composite, the percolated metallic NM network increases the effective conductive cross-section with thickness, resulting in $R_{\text{int}}^{3\text{D}} \propto 1/t$. Within the experimentally accessible thickness range ($\gg \lambda_{\text{m}}$), the effective SSE is not limited by interfacial magnon diffusion and therefore does not exhibit the saturation characteristic of two-dimensional systems. Consequently,

$$P_{\text{max}}^{3\text{D}}(t) \propto t, \quad (6)$$

leading to a non-saturating increase of output power with increasing thickness. A detailed derivation of these scaling relations, including expressions in terms of experimentally measurable parameters, is provided in Supplementary Note 5.

For the three-dimensional composite, experimentally measured transport parameters from the 30-300 and 30-RT samples were used for the scaling analysis, including $\kappa$ = 1.87 W/mK, $\sigma = 2.31 \times 10^3$ S/m, and $S_{SSE}$ = 14.8 nV/K for the 30-RT sample. For the quasi-two-dimensional reference systems, representative material parameters reported for polycrystalline $NiFe_2O_4$/Pt[21], single-crystalline YIG/Pt[69], and epitaxial YIG films /$WSe_2$/Pt[34] heterostructures were adopted, using literature values of $S_{SSE}$, $\kappa$ of the FM layer and $\sigma$ of the NM layer. Using these parameters (see Supplementary Table 1), Fig. 5 presents $P_{max}$ as a function of thickness under identical device area (10 mm × 10 mm) and heat-flux ($j_Q = 1.5 \times 10^4$ W/m$^2$) boundary conditions. For the three-dimensional composite, the scaling analysis is shown only in the regime where the thickness exceeds both $\lambda_m$ and the experimentally accessible range (> 0.1 mm), corresponding to the volumetric scaling regime. The quasi-three-dimensional composite architecture exhibits an approximately linear increase in output power with thickness, whereas conventional quasi-two-dimensional devices exhibit saturation behavior governed by $\lambda_m$. This contrast demonstrates that the present architecture removes the intrinsic interfacial thickness limitation characteristic of thin-film SSE systems and establishes a bulk-operating scaling regime enabled by volumetric spin-to-charge conversion. The dashed lines in Fig. 5 represent projected scaling behavior assuming higher intrinsic $S_{SSE}$ comparable to those reported in optimized thin-film systems (on the order of 1 μV/K), while maintaining the same three-dimensional architecture and boundary conditions. Because the maximum output power in the volumetric regime follows $P_{max}^{3D} \propto S_{SSE}^2 t$, increasing the intrinsic $S_{SSE}$ enhances the slope of the linear thickness scaling without altering the linear thickness dependence.

## Discussion

In summary, we have demonstrated the observation of trans-scale SSE using YIG-Pt bulk composite materials. The composites were fabricated via a dynamic powder sputtering method followed by low-temperature sintering. These bulk materials exhibit clear transverse thermoelectric signals at the bulk scale, consistent with volumetric spin-to-charge conversion. Through rigorous signal and material characterization, we confirm that the observed signals originate purely from the SSE of YIG, with negligible parasitic contributions.

We believe this study represents a key step toward the development of practical SSE-based thermoelectric devices by bridging the technical gap between conventional nanoscale SSE effects, which are limited by the spin and magnon diffusion lengths, and bulk device platforms. From the intrinsic (materials) perspective, further improvement in output power can be achieved by optimizing FM/NM interfaces[21,22], for example by inserting topological materials[34,70] or organic semiconductors[71] at the interface, as well as engineering magnon transport in YIG independently from phonons[16,21]. In addition, replacing YIG with alternative high-performance FMI with lower Gilbert damping constants and longer magnon diffusion lengths, and adopting NM materials with a giant inverse spin Hall effect, such as topological insulators and Weyl semimetals[72–77], could substantially increase conversion efficiency at bulk scale. These NM materials offer long spin diffusion lengths (up to 2 μm at low temperature[78,79] and even at room temperature[80]) and spin Hall conductivities that exceed those of Pt by orders of magnitude, providing a direct route to boost spin Seebeck coefficient and the resulting output power. Many of these strategies have already been demonstrated independently in thin-film spintronics and spin caloritronics, where significant performance improvements compared to Pt-based systems have been achieved. Translating these advances into bulk architectures could therefore accelerate the development of practical SSE-based energy harvesters. Moreover, hybridization with other transverse thermoelectric phenomena, such as the ANE[7,10,43,81], which has recently shown remarkable performance enhancements in topological materials[82–86] and structurally heterogeneous systems[45,87–89], may provide an additional route for enhancing bulk-scale performance. Hybrid transverse thermoelectric responses combining SSE and ANE have been reported in YIG/FMM bilayer systems, for example using Ni as the ferromagnetic metal[9], where anomalous Nernst response and spin-to-charge conversion coexist within the same device geometry. In this context, employing FMMs that exhibit both a sizable ANE and efficient spin-to-charge conversion, including anomalous inverse ISHE[90], could enable constructive coexistence of ANE- and SSE-driven transverse voltages within the same device geometry. Because both effects share identical symmetry with respect to magnetization and thermal gradient, their contributions may, in principle, superpose and yield synergistic enhancement when their signs are aligned.

From the extrinsic (structural) perspective, our symmetry analysis in Equation (3) reveals that the trans-scale SSE amplitude in the present composites contains a geometric reduction factor of 1/3 arising from the statistically isotropic distribution of local FMI/NM interface normals. Structural engineering that induces anisotropic or partially aligned interface orientations, for example, via controlled grain-shape anisotropy, magnetic-field-assisted forming, or directional sintering, can bias the distribution of interface normals and increase the geometric contribution toward the planar-film limit (i.e., the reduction factor is 1). This identifies structural orientation control as an additional, previously unexplored degree of freedom for amplifying trans-scale SSE signals in 3D architectures.

Altogether, the demonstration of trans-scale SSE in a bulk composite platform establishes a previously inaccessible regime of transverse thermoelectric conversion that extends well beyond the constraints of conventional thin-film architectures. By enabling volumetric spin-current generation and macroscopic ISHE conversion within a mechanically robust 3D network, this work provides a practical and scalable route for implementing spin-based thermoelectric functionality at bulk scale. Unlike interfacial thin-film SSE systems, the present architecture supports bulk-scalable geometries in which output power increases with device thickness without interfacial saturation, thereby expanding the accessible design space for transverse thermoelectric energy conversion. These results position trans-scale SSE composites as a promising platform for next-generation transverse thermoelectrics based on spin currents.

## Methods

### *Sample preparation*

The Pt-coated YIG powders were prepared using a customized dynamic sputtering system for powder coating (SUGA Co., Ltd., SSP-1500B) with a base pressure of less than 5 $\times 10^{-4}$ Pa. The initial YIG powders were prepared by crushing YIG single crystals (Ferrisphere Inc.). The sputtering system vibrated the powders at a frequency of 22.5 Hz and rotated the powders at approximately 17 rpm to stir them. Pt was sputtered from a 99.99%-purity Pt target with an Ar gas flow of 6.0 sccm at room temperature. The plasma power was set to 50 W. The 5-nm Pt sample shown in Fig. 2 was deposited for 30 min, and the Pt thickness of the main samples (15 nm and 30 nm) were labeled based on the corresponding deposition durations.

The Pt-coated YIG powders were then sintered into bulk form by hot pressing under two low-temperature, high-pressure conditions: (1) 300 °C at 350 MPa and (2) room temperature at 500 MPa. Sintering was performed using a hot-pressing apparatus (AS ONE, H400-15) in air, employing a tungsten carbide die and punches with a diameter of 10 mm. The samples were cut using a diamond wire saw (Diamond WireTec, DWS.100) for subsequent materials characterization. All properties reported in this study were measured from samples prepared in a single batch.

### *Structural characterization*

Structural analysis was performed using a Rigaku MiniFlex600 X-ray diffractometer with Cr-Kα radiation (wavelength = 0.22897 nm). The diffraction angles were converted to commonly used Cu-Kα equivalent (wavelength = 0.15406 nm) for presentation in

Supplementary Fig. 3. The X-ray tube was operated at 40 kV and 15 mA.

Microstructural observations were performed using a Carl Zeiss CrossBeam 1540EsB scanning electron microscope (SEM) equipped with energy-dispersive X-ray spectroscopy (EDS). Transmission electron microscopy (TEM) was carried out with a Titan G2 80–200 microscope equipped with a probe aberration corrector. TEM specimens were prepared using the lift-out technique with an FEI Helios 5UX dual-beam focused ion beam system.

### *Measurement of thermal and electrical conductivities*

The electrical conductivity ($\sigma$) was measured using a standard four-probe method with a current source (Keithley, 2450) and a nanovoltmeter (Keithley, 2182a). A linear voltage-current response was confirmed by measuring 9 points with reverse polarities of the current.

The thermal conductivity ($\kappa$) was estimated by multiplying thermal diffusivity, heat capacity, and density. Owing to the statistically random orientation of Pt-coated YIG particles in the sintered composite, the macroscopic thermal and spin transport properties are treated as effectively isotropic in the device-level analysis. The thermal diffusivity was obtained using a laser flash method (LINSEIS, LFA1000), specific heat was measured by DSC (Rigaku, Thermo plus EVO2), and density was calculated from the sample mass and volume at room temperature. The DSC measurement was conducted with an aluminum pan over a temperature range of 5 ℃ to 45 ℃ with a ramp rate of 1 ℃/min. Alumina ($Al_2O_3$) powder was used as a reference for the DSC measurement.

### *Measurement of the spin Seebeck effect*

The SSE was measured in both $H_z$-$\nabla_x T$ and $H_x$-$\nabla_z T$ configurations. For $H_z$-$\nabla_x T$ configuration, two Cu wires were attached along the $y$-direction using silver epoxy and the sample was sandwiched between two thermally conductive, electrically insulating sapphire blocks serving as heat baths. Thermal grease was used at the interfaces to ensure a uniform temperature gradient. The $x$-directional temperature gradient applied was controlled using a PID-based temperature controller with measured temperature differences using differential thermocouples. It is worth noting that the measured temperature difference includes not only the drop across the sample but also the interfacial drops. This leads to an overestimation of the applied gradient across the samples and thus an underestimation of the reported spin Seebeck coefficients. Although this effect has been reported to be appreciable even at room temperature for YIG single crystals (despite the use of thermal grease)[91], we expect the error here to be smaller because the thermal conductivity of our samples is lower (1.46–1.87 W/mK), i.e., about 18–27% of that of single-crystalline YIG (~7–8 W/mK[46,54]). Consequently, a larger fraction of the total temperature drop occurs within the sample rather than at the interfaces. A current source (ADVANTEST, R6243) and a multimeter (Keithley, 2000) were used for thermal gradient control. The voltage was measured using a nanovoltmeter (Keithley, 2182a) while applying magnetic fields with an electromagnet. The $H_x$-$\nabla_z T$ configuration was conducted in a similar manner, with the 30-RT sample bridged by two Cu blocks to apply a $z$-directional temperature gradient, as described in [87,92]. Adhesive thermal tapes (3M, VHR0601-03) were used to electrically insulate the sample from the Cu blocks.

***Magnetic property analysis***

The magnetic properties of the samples were investigated using a SQUID

magnetometer (Quantum Design, MPMS). The measurements were conducted at $T$ = 300 K with a sweep rate of 100 Oe/s. For the fine-scan measurement in Supplementary Fig. 6c, the magnetization was recorded at each field after stabilizing the magnetic field.

***Measurement of the Hall effect***

The Hall effect shown in Fig. 3e was measured under $H_x$ and $I_z$ (same as the $H_x$-$\nabla_z T$ configuration) at room temperature using the DC resistivity option of a He cryostat (Cryogenic Limited, CFMS). Two electrodes were attached at the sample ends with silver epoxy to apply a current using a current source (Keithley, 2450) and two additional electrodes were attached orthogonally to capture the Hall voltage using a nanovoltmeter (Keithley, 2182a). The input current was 100 mA. The measurement was conducted over a magnetic field range of $\pm$ 10 kOe, consistent with the $H_x$-$\nabla_z T$ SSE measurement, using a sweep rate of 50 Oe/s. The data in Fig. 3e were symmetrized to remove the magnetic-field-even dependent components (e.g., magnetoresistance), following the same procedure used for the SSE data.

Data availability

All data that support the findings of this study are available within the article and Supplementary Information. Source data underlying the plots in the main and supplementary figures are provided with this paper as a Source Data file.

## Acknowledgements

We thank W. Zhou for helpful discussions and experimental support during the early stage of this work.

## Funding

This work was supported by ERATO "Magnetic Thermal Management Materials" (grant no. JPMJER2201) from JST, Japan.

Author Contributions

S.J.P. fabricated the Pt-coated YIG powders, measured and analyzed the VSM, XRD, electrical and Hall conductivities, thermal diffusivity, DSC and SSE data. K.H. synthesized the bulk composites at low-temperature and high-pressure conditions. H.S.A. performed the SEM and TEM measurements. F.A. and T.H. assisted with the SSE and thermal diffusivity measurements, respectively. K.U. conceived and supervised the project. S.J.P. wrote the manuscript with input from all authors.

Competing interests

The authors declare no competing interests.

Figures

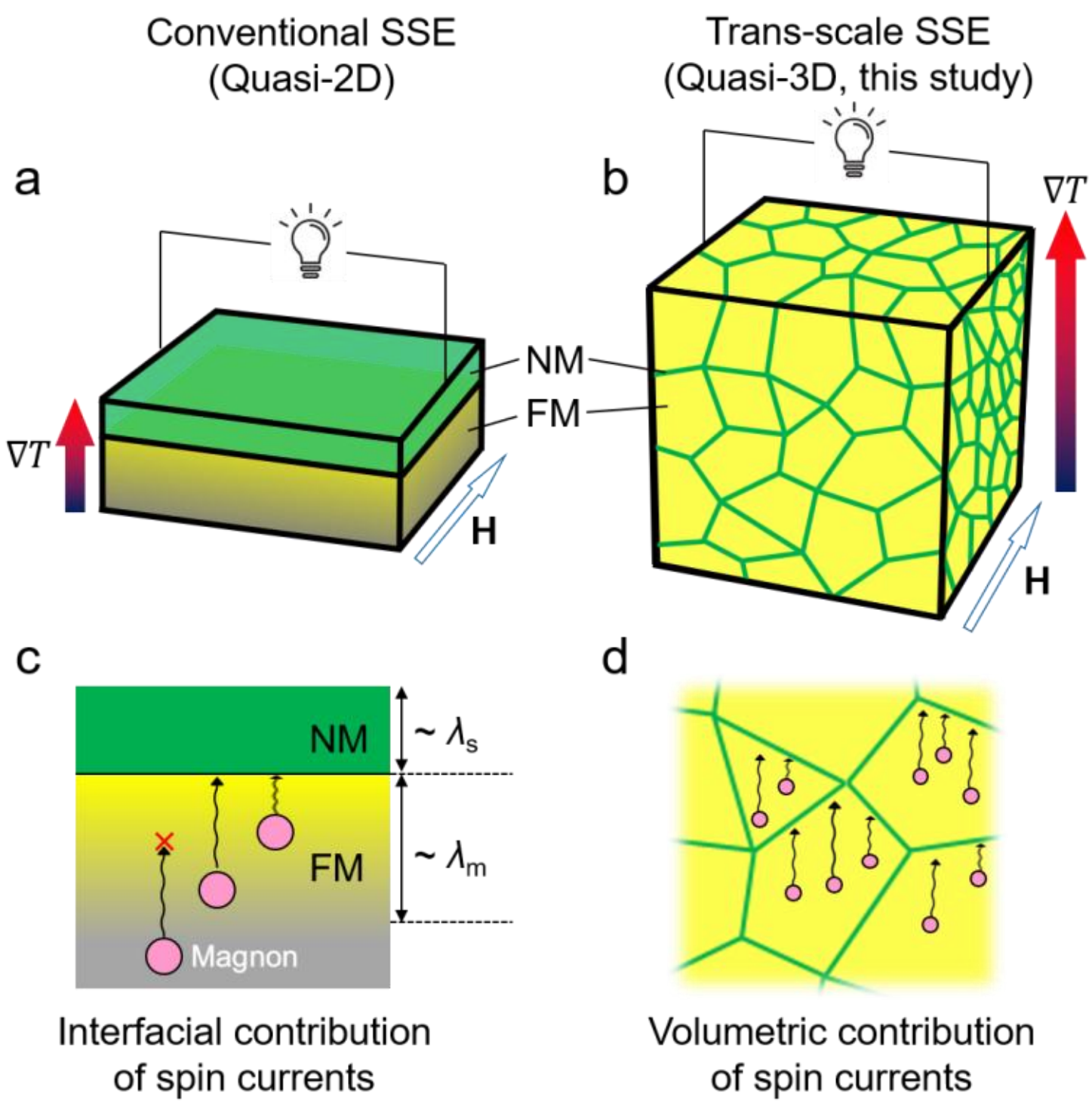

**Fig. 1| Conceptual comparison between conventional thin-film-based spin Seebeck effect (SSE) and the proposed trans-scale SSE.** (a) Schematic of a typical quasi-two-dimensional SSE device composed of a ferro(i)magnetic material (FM) and a normal metal (NM) layer. The output electric field is generated orthogonal to the applied thermal gradient ($\nabla T$) and magnetic field (**H**). (b) Schematic of a trans-scale SSE device, realized via a bulk composite of NM-coated FM domains. (c) In the conventional SSE device, the device length scale is determined by spin diffusion length of NM $\lambda_s$ and magnon diffusion length of FM $\lambda_m$. Here, only spin currents and magnons that reach the FM/NM interface contribute to SSE, limiting device performance. (d) In the trans-scale SSE, distributed NM channels allow efficient local conversion of magnons throughout the FM volume, enhancing effective spin current utilization.

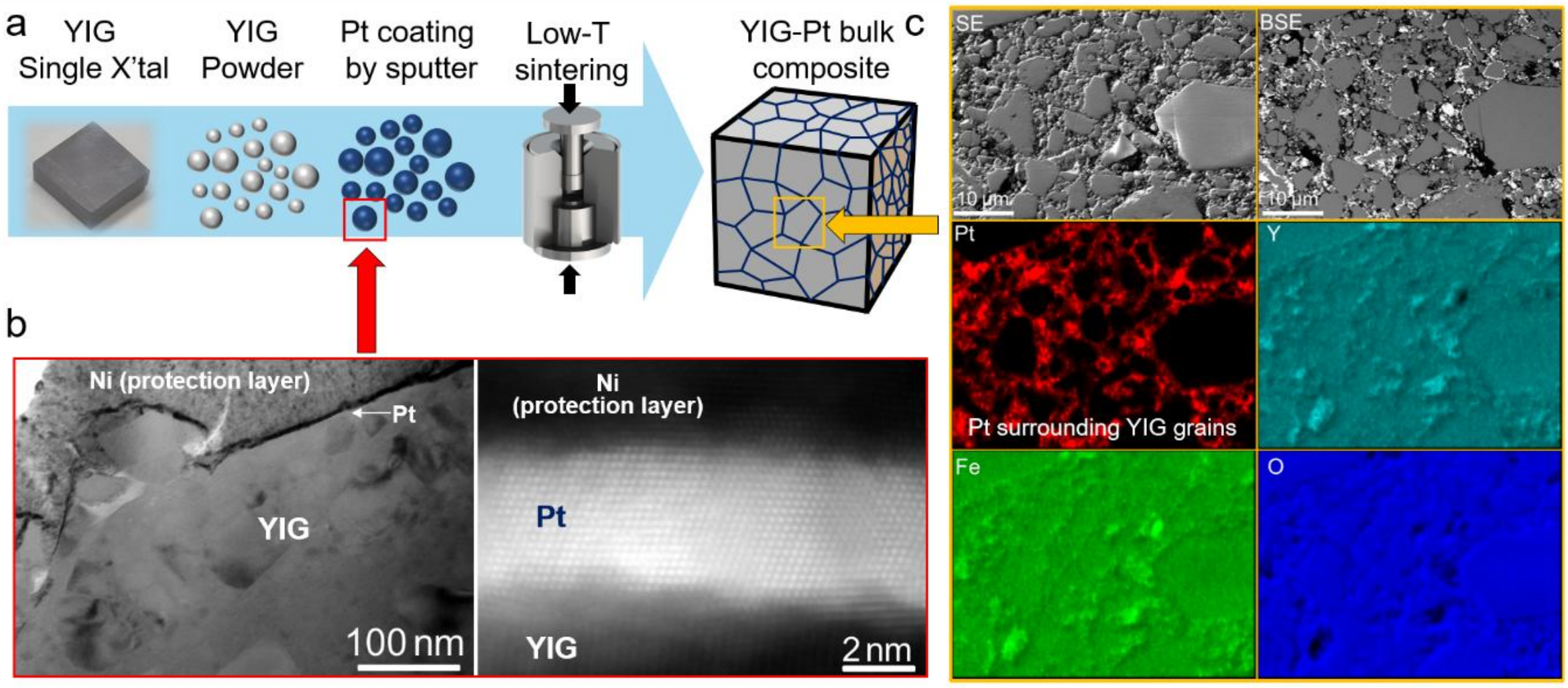

**Fig. 2| Fabrication and characterization of YIG-Pt composite materials for trans-scale spin Seebeck effect.** (a) Schematic illustration of the fabrication process. YIG powders are coated with Pt using dynamic powder sputtering, followed by low-temperature sintering to produce bulk composites. (b) TEM images showing uniform nm-scale Pt coating on the surface of YIG powders. (c) Microstructural and compositional characterization of the bulk composite. SEM images (SE and BSE) reveal grain size distribution. EDX elemental mappings confirm uniform distribution of Y, Fe, and O, throughout the composite matrix and Pt surrounding YIG grains.

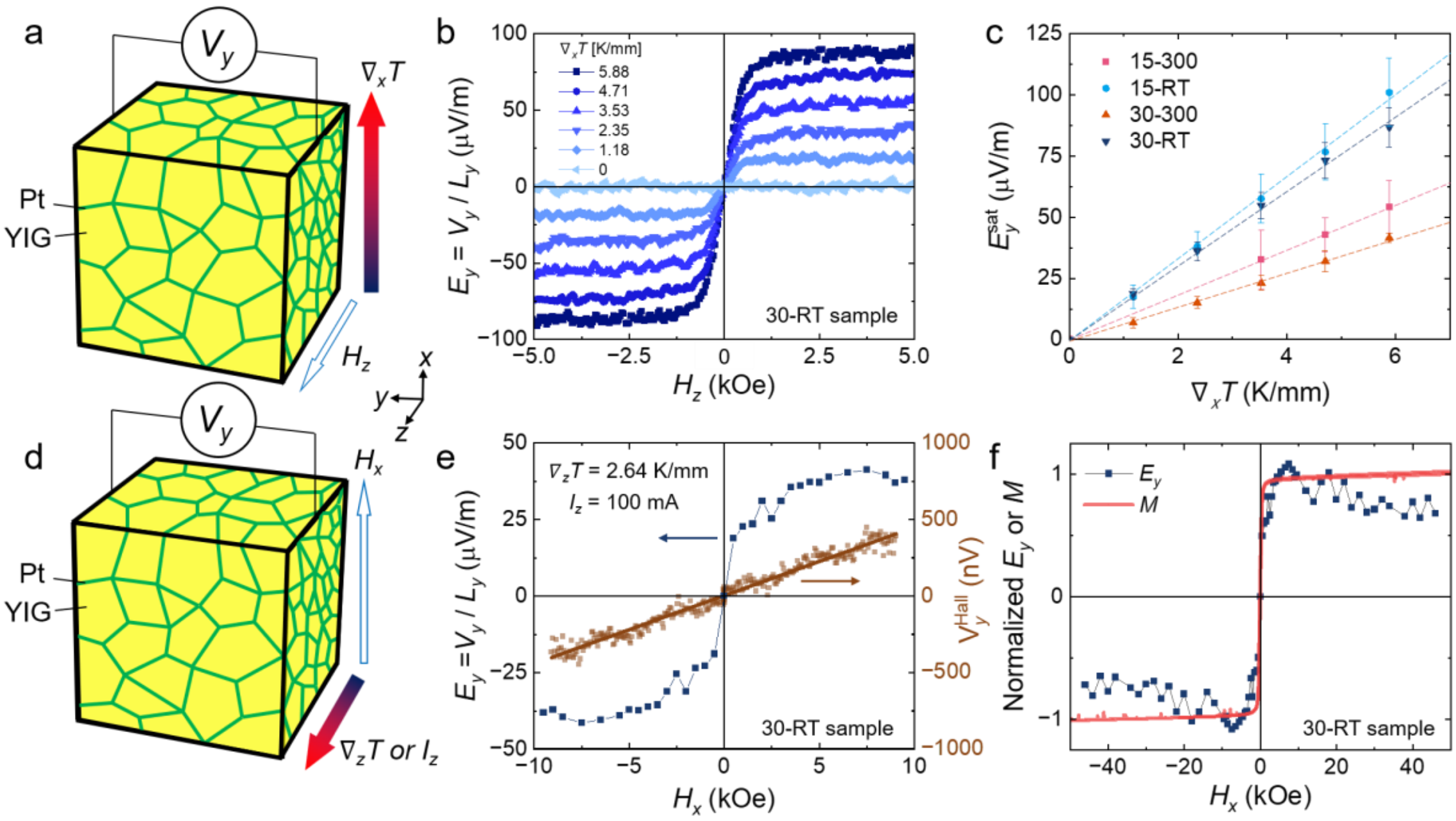

**Fig. 3| Observation of the trans-scale spin Seebeck effect.** (a) Schematic illustrations of $H_z$-$\nabla_x T$ configuration. (b) Measured transverse electric field ($E_y$) under $\nabla T$ in $H_z$- $\nabla_x T$ configuration as a function of $H$. The subscripts $x$, $y$, and $z$ indicate the corresponding directions of the applied temperature gradient ($\nabla T$), charge current ($I$), and magnetic field ($H$). (c) Saturated electric field ($E_y^{\mathrm{Sat}}$) plotted as a function of $\nabla_x T$. Error bars in (c) represent the standard deviation. (d) Schematic illustrations of $H_x$- $\nabla_z T$ configuration. (e) Measured transverse electric field ($E_y$) under $\nabla T$ in $H_x$-$\nabla_z T$ configuration. The measured Hall voltage ($V_y^{\mathrm{Hall}}$) is also presented for reference, along with a linear fitting curve. (f) Normalized $E_y$ and Magnetization ($M$) of the 30-RT sample under high $H$. Detailed $M$-$H$ curves for the other samples are provided in Supplementary Figure 6.

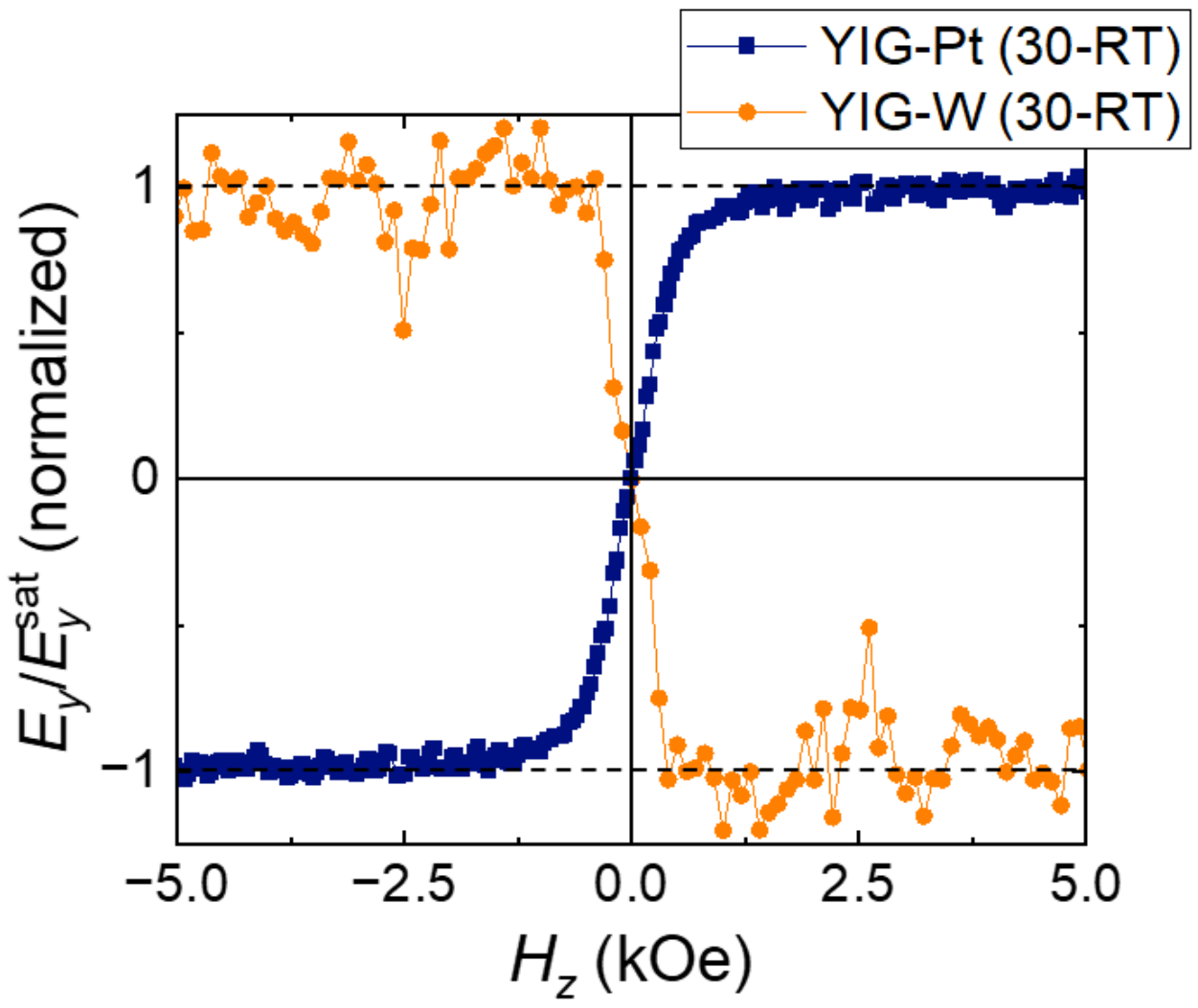

**Fig. 4| Normalized spin Seebeck voltage for YIG–Pt and YIG–W bulk composites.** The two samples exhibit opposite signs of transverse thermoelectric voltage, reflecting the opposite signs of the spin Hall angles of the nonmagnetic metals, Pt (positive) and W (negative). The signal in the YIG–W composite appears noisier due to the large electrical resistance of the non-optimized composite. See Supplementary Note 4 for fabrication details and discussion.

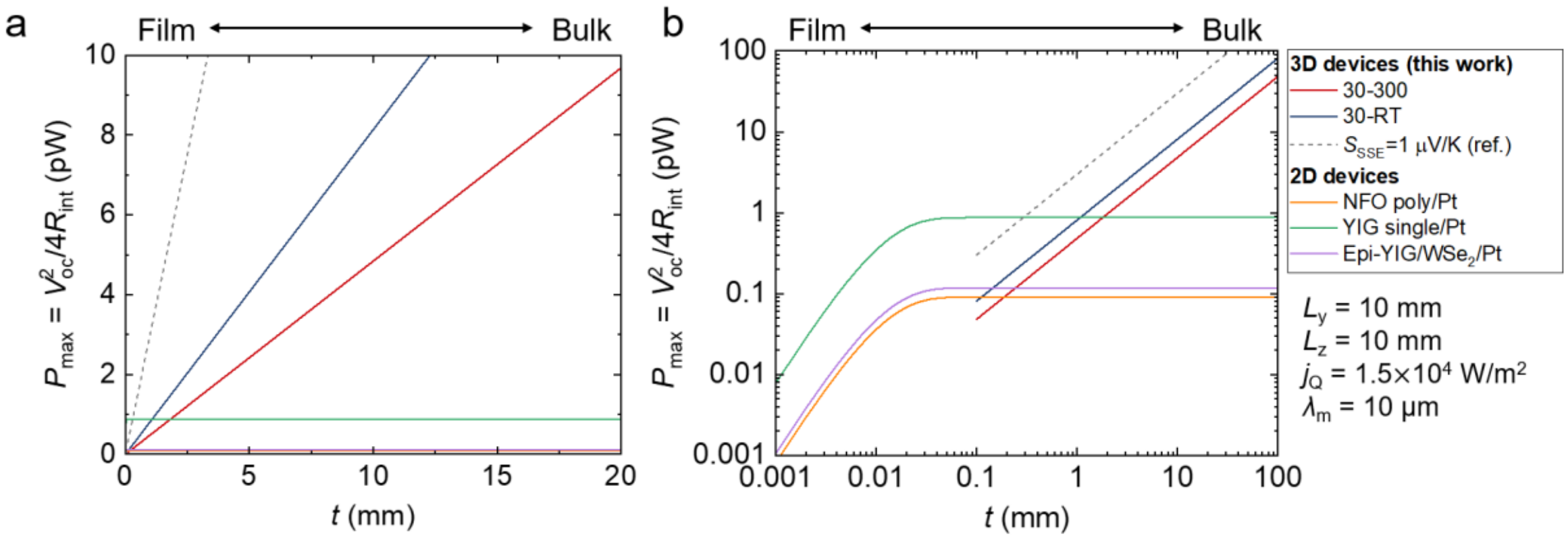

**Fig. 5| Thickness-dependent scaling of maximum output power in transverse spin Seebeck devices.** (a) Linear-scale plot and (b) logarithmic-scale plot of the maximum output power $P_{\max}$ as a function of ferromagnetic layer thickness. $P_{\max}$ was calculated under identical device area ($L_y = L_z = 10$ mm) and heat-flux ($j_Q = 1.5\times10^4$ W/m$^2$) conditions using $P_{\max} = V_{oc}^2/4R_{int}$, where $V_{oc}$ is the open circuit voltage and $R_{int}$ is the internal resistance. A representative magnon diffusion length of $\lambda_m = 10$ μm was used for the 2D reference calculations. The 2D reference systems were modeled using representative material parameters from polycrystalline $NiFe_2O_4$ (NFO poly)/Pt[21], single-crystalline YIG (YIG single)/Pt[69], and epitaxial YIG films (Epi-YIG)/$WSe_2$/Pt[34] heterostructures reported in the literature. For the 3D architecture, the thickness-dependent scaling is shown only for thicknesses exceeding both $\lambda_m$ and the experimentally accessible range (> 0.1 mm), thereby representing the volumetric scaling regime. Detailed assumptions and parameter values are provided in Supplementary Note 5.

# Supplementary Information of

# Trans-scale spin Seebeck effect in nanostructured bulk composites based on magnetic insulator

Sang J. Park[1,*], Keisuke Hirata[2,3], Hossein Sepehri-Amin[1], Fuyuki Ando[1], Takamasa Hirai[1], and Ken-ichi Uchida[1,3,*]

[1] National Institute for Materials Science, Tsukuba 305-0047, Japan

[2] Toyota Technological Institute, Nagoya 468-8511, Japan

[3] Department of Advanced Materials Science, Graduate School of Frontier Sciences, The University of Tokyo, Kashiwa 277-8561, Japan

*Correspondence to: PARK.SangJun@nims.go.jp (S.J.P.); UCHIDA.Kenichi@nims.go.jp (K.U.)

## Supplementary Notes

### Supplementary Note 1. Symmetry-based derivation of macroscopic ISHE field

Here we provide a brief derivation of Equations (2)–(3) in the main text, showing how a finite macroscopic inverse spin Hall effect (ISHE) field arises in a ferromagnetic material (FM)/normal metal (NM) composite with randomly oriented interface facets.

### Local ISHE response for an arbitrary interface

For a local FM/NM interface patch with unit normal $\hat{n}$, the spin Seebeck effect (SSE) pumps a spin current perpendicular to the interface. Its local density can be expressed as

$$\mathbf{j}_s = j_0(\hat{n} \cdot \nabla T)\hat{n} \tag{S1}$$

where $\nabla T$ is the applied temperature gradient and $(\hat{n} \cdot \nabla T)$ selects its normal component. The ISHE converts this spin current into a transverse electric field,

$$\mathbf{E}_{\mathrm{ISHE}}(\hat{n}) \propto \mathbf{j}_s(\hat{n}) \times \hat{m} = (\hat{n} \cdot \nabla T)(\hat{n} \times \hat{m}) \tag{S2}$$

with the magnetization direction $\hat{m}$. Equation (S2) reproduces Eq. (2) in the main text. Under inversion of the interface normal ($\hat{n} \rightarrow -\hat{n}$), both factors in Eq. (S2) change sign,

$$(\hat{n} \cdot \nabla T) \rightarrow -(\hat{n} \cdot \nabla T), \qquad (\hat{n} \times \hat{m}) \rightarrow -(\hat{n} \times \hat{m}) \tag{S3}$$

so their product, and thus the local ISHE field, remains unchanged:

$$\mathbf{E}_{\mathrm{ISHE}}(-\hat{n}) = \mathbf{E}_{\mathrm{ISHE}}(\hat{n}). \tag{S4}$$

This invariance explains why concave, convex, rough, or oppositely oriented interface facets contribute with the same sign.

### Orientational average for an isotropic interface network

In a polycrystalline bulk composite, the FM/NM interfaces are oriented in all directions. For such an isotropic distribution of interface normals, the average products of their components along two spatial directions (i.e., $n_i n_j$) must reflect this symmetry. As a result, the orientational average takes the rotationally symmetric form

$$\langle n_i n_j \rangle = \frac{1}{3}\delta_{ij} \tag{S5}$$

meaning that the components of $\hat{n}$ are equally sampled in all directions. Averaging the local response in Equation (S2) using Equation (S5) yields

$$\langle \mathbf{E}_{\mathrm{ISHE}} \rangle \propto \frac{1}{3}(\nabla T \times \hat{m}) \tag{S6}$$

which is Equation (3) in the main text. This result shows that even in a geometrically complex FM/NM composite, the macroscopic SSE preserves the usual ISHE symmetry $\nabla T \times \hat{m}$ while being reduced only by a geometric factor of 1/3 due to isotropic averaging.

**Supplementary Note 2. Discussion on low-temperature sintering of YIG-Pt composites**

Here, we discuss the low-temperature sintering of YIG-Pt composites. As stated in the main text, oxide YIG powders require high temperatures for sintering, typically over 900 ℃ due to their thermodynamically stable crystal structures and strong ionic and covalent bonding[1]. However, such high-temperature processing can lead to degradation of the thin Pt layers, including increased surface roughness, dewetting, and recrystallization into nanoparticles to minimize surface free energy. For example, Golosov et al.[2] reported the formation of hillocks several hundred nanometers high upon annealing sputter-deposited Pt-based multilayers at temperatures as low as 400 ℃, which correlated with changes in electrical resistivity. Similarly, Sui et al.[3] observed that annealing 10-nm-thick sputter-deposited Pt layers at 550 ℃ led to agglomerated nanostructures with significantly increased surface roughness and changed optical properties.

At the initial stage of the study, we performed high-temperature sintering of YIG-Pt powders at 800 °C, a temperature slightly lower than that reported in [1], to obtain high-density nanostructured composite samples. Although the resulting pellet exhibited a well-densified structure, it showed extremely high electrical resistance, exceeding the measurement range of our system (> MΩ), consistent with the expected degradation of thin Pt layer quality.

We therefore fabricated the YIG-Pt bulk pellets at low temperatures (300 °C and room temperature) under high pressure conditions, as described in the main text (Methods). The density of the samples reached up to 75% (30-RT sample, Supplementary Fig. 4d) relative to the YIG single crystal (5.11 g/cm$^3$). The low-temperature sintering was enabled by the surface coating of ductile metallic Pt layers. As a result, the Pt-coated YIG powders were more favorably pelletized through the Pt channels, providing mechanical adhesion between the powders. We also pelletized the YIG powders using the same conditions, intended as a control sample. However, we obtained a sample with poor mechanical robustness, exhibiting many macroscale cracks, suggesting the critical role of Pt coating for sintering.

In summary, the Pt coating of oxide YIG powders provides additional mechanical robustness to the system, allowing for low-temperature sintering.

**Supplementary Note 3. Phase analysis based on XRD**

We discuss potential structural changes during the deposition and sintering processes. $\theta$-2$\theta$ powder XRD measurements were conducted using a Cr-Kα beam (wavelength = 0.22897 nm, Methods). The 2$\theta$ values in Supplementary Fig. 3 were converted to equivalent values based on the commonly used Cu-Kα radiation (wavelength = 0.15406 nm) for easier comparison.

The XRD pattern of the initial YIG powders, obtained by crushing YIG single crystals, is shown in Supplementary Fig. 3a. The predominant phase was identified as YIG, in agreement with the reference peaks of YIG (PDF#43-0507), without evidence of preferred crystal orientation. Only negligibly small peaks corresponding to $Y_2O_3$ (PDF#43-0661) were observed near 2$\theta$ = 29°. The XRD data from the main composite samples (15-300, 15-RT, 30-300, and 30-RT) consistently exhibited this small peak. The only noticeable difference between the composite samples and the initial YIG powder was the presence of broad Pt peaks (PDF#04-0802) near 2$\theta$ of 40° and 46°, indicative of low-crystallinity Pt formed on YIG. These results suggest that neither the room-temperature Pt sputtering nor the low-temperature sintering introduced additional secondary phases detectable by XRD.

To further exclude the formation of secondary magnetic phases, we replotted the XRD data (Supplementary Fig. 3a) with reference peaks of Fe- and Pt-based ferromagnetic materials ($Fe_3O_4$: PDF#19-0629 and FePt: PDF#43-1359), as shown in Supplementary Fig. 3b. These materials are known to contribute to transverse thermoelectric signals via the anomalous Nernst effect, which can be captured in both in-plane (Fig. 3a) and out-of-plane (Fig. 3d) magnetization configurations in the main text. However, no identifiable peaks corresponding to these phases were observed, confirming the absence of contamination in the observed trans-scale SSE signals shown in Fig. 3 of the main text.

**Supplementary Note 4. Observation of trans-scale SSE in YIG–W bulk composite**

To unambiguously verify the origin of the transverse thermoelectric voltage observed in the YIG–Pt composites, we performed a control experiment by replacing Pt with W, which has a negative spin Hall angle. The YIG–W composite was fabricated using the same procedure as the YIG–Pt samples (specifically the 30-RT condition). W was deposited onto the surface of YIG powders using the powder sputtering system and subsequently consolidated by high-pressure pressing at 500 MPa, identical to the procedure used for the Pt-based composite. The resulting YIG–W composite exhibited higher mechanical brittleness than the YIG–Pt samples, possibly due to the non-optimized fabrication conditions and/or partial oxidation of the W shells. The electrical resistance of the sample exceeded 10 kΩ, more than four orders of magnitude higher than that of the YIG–Pt composite with similar dimensions, suggesting that the W network was not fully continuous throughout the bulk. Further optimization may be required to obtain mechanically robust, electrically connected YIG–W composites. Due to the mechanical brittleness of the YIG–W composite, the samples could not be cut into well-defined rectangular geometries. Measurements were therefore performed using irregular pellet-like pieces, which makes absolute normalization by geometric dimensions challenging. To allow a meaningful comparison with the YIG–Pt results, the data presented in Fig. 4 of the main text are normalized to their maximum values. Despite these non-idealities, we clearly observed a transverse thermoelectric voltage with the opposite sign compared with the YIG–Pt composite, as shown in Fig. 4 in the main text. Although the signal is noisier possibly due to the high resistance of the sample, the sign reversal is evident and exceeds the fluctuation level. This behavior is fully consistent with the expected sign change of the spin Hall angle when replacing Pt (positive) with W (negative), thereby providing direct and independent evidence that the transverse thermoelectric voltage originates from magnon-driven thermal spin pumping in YIG and the spin-charge conversion via the ISHE in the NM.

## Supplementary Note 5. Thickness-dependent scaling of output power in 2D and 3D spin Seebeck devices

### General expression for output power

The transverse electric field generated by the SSE is expressed as

$$E_y = S_{\mathrm{SSE}} \nabla T, \tag{S7}$$

where $S_{\mathrm{SSE}}$ is the spin Seebeck coefficient and $\nabla T$ is the applied temperature gradient. The open-circuit voltage measured along the device length $L_y$ is therefore

$$V_{\mathrm{oc}} = S_{\mathrm{SSE}} \nabla T L_y. \tag{S8}$$

When the device is connected to an external load resistance $R_L$, the output power is

$$P(R_L) = \frac{V_{\mathrm{oc}}^2 R_{\mathrm{L}}}{(R_{\mathrm{int}}+R_{\mathrm{L}})^2}, \tag{S9}$$

where $R_{\mathrm{int}}$ is the internal resistance of the device. Maximum power transfer occurs when $R_{\mathrm{L}} = R_{\mathrm{int}}$, yielding

$$P_{\mathrm{max}} = \frac{V_{\mathrm{oc}}^2}{4R_{\mathrm{int}}}. \tag{S10}$$

This expression forms the basis for the thickness-dependent scaling analysis presented below.

### Thickness dependence in conventional quasi-two-dimensional SSE systems

In conventional quasi-two-dimensional SSE devices, spin-current generation and spin-to-charge conversion are interfacial processes. The $S_{\mathrm{SSE}}$ therefore depends on the magnetic layer thickness $t$ as

$$S_{\mathrm{SSE}}(t) = S_0(1 - e^{-t/\lambda_m}), \tag{S11}$$

where $S_0$ is the bulk spin Seebeck coefficient and $\lambda_{\mathrm{m}}$ is the magnon diffusion length.

The open-circuit voltage becomes

$$V_{\mathrm{oc}}^{\mathrm{2D}}(t) = S_0(1 - e^{-t/\lambda_m}) \nabla T L_y. \tag{S12}$$

Under fixed normal-metal geometry, the internal resistance is approximately independent of the magnetic layer thickness ($R_{\mathrm{int}}^{\mathrm{2D}} \approx$ const.).

Accordingly, the maximum output power follows

$$P_{\mathrm{max}}^{\mathrm{2D}}(t) = \frac{\left[S_0\left(1-e^{-\frac{t}{\lambda_m}}\right)\nabla T L_y\right]^2}{4R_{\mathrm{int}}^{\mathrm{2D}}} \propto (1 - e^{-t/\lambda_m})^2. \tag{S13}$$

As $t$ exceeds $\lambda_{\mathrm{m}}$, both the voltage and the output power approach saturation. In the present scaling analysis, we adopt a representative magnon diffusion length of $\lambda_{\mathrm{m}} \sim 10$ μm at room temperature as reported in the literature[4]. We note that additional magnon energy-relaxation mechanisms, which have been discussed at shorter length scales on the order of ~250 nm, are not included in this simplified model. Since the thickness range considered in this work significantly exceeds the sub-micron regime, the diffusion-limited description provides an appropriate approximation for the purpose of architectural scaling comparison.

**Thickness dependence in quasi-three-dimensional volumetric SSE composites**

In the present three-dimensional composite architecture, spin-to-charge conversion occurs throughout the bulk volume via a percolated metallic network.

Under fixed lateral dimensions $L_y$ and $L_z$, increasing thickness $t$ increases the effective conductive cross-section of the interconnected metallic pathways. Using $\sigma$, the internal resistance scales as

$$R_{\mathrm{int}}^{\mathrm{3D}} \approx \frac{1}{\sigma}\frac{L_y}{L_z t} \propto \frac{1}{t}. \tag{S14}$$

Within the experimentally relevant thickness range (> 0.1 mm), the $S_{\mathrm{SSE}}$ does not exhibit the interfacial saturation characteristic of quasi-two-dimensional systems. The open-circuit voltage can therefore be expressed as

$$V_{\mathrm{oc}}^{\mathrm{3D}} = S_{\mathrm{SSE}}^{\mathrm{3D}} \nabla T L_y, \tag{S15}$$

where $S_{\mathrm{SSE}}^{\mathrm{3D}}$ is approximately thickness-independent in the absence of interfacial limitation.

Substituting into the general power expression yields

$$P_{\mathrm{max}}^{\mathrm{3D}}(t) = \frac{[S_{\mathrm{SSE}}^{\mathrm{3D}} \nabla T L_y]^2}{4R_{\mathrm{int}}^{\mathrm{3D}}} \propto t. \tag{S16}$$

Thus, in contrast to the saturation behavior of 2D interfacial SSE systems, the three-dimensional composite architecture exhibits a non-saturating power scaling with thickness within the accessible regime.

**Architectural origin of the scaling difference**

The key distinction between the two architectures arises from the spatial distribution of spin-to-charge conversion:

- In two-dimensional systems, conversion is confined to a thin interfacial region, leading to intrinsic thickness saturation governed by the magnon diffusion length.
- In the three-dimensional composite, conversion occurs throughout the bulk volume, while the internal resistance decreases with increasing conductive cross-section.

This difference in conversion topology results in fundamentally different thickness-scaling behaviors of output power.

## Supplementary Figures

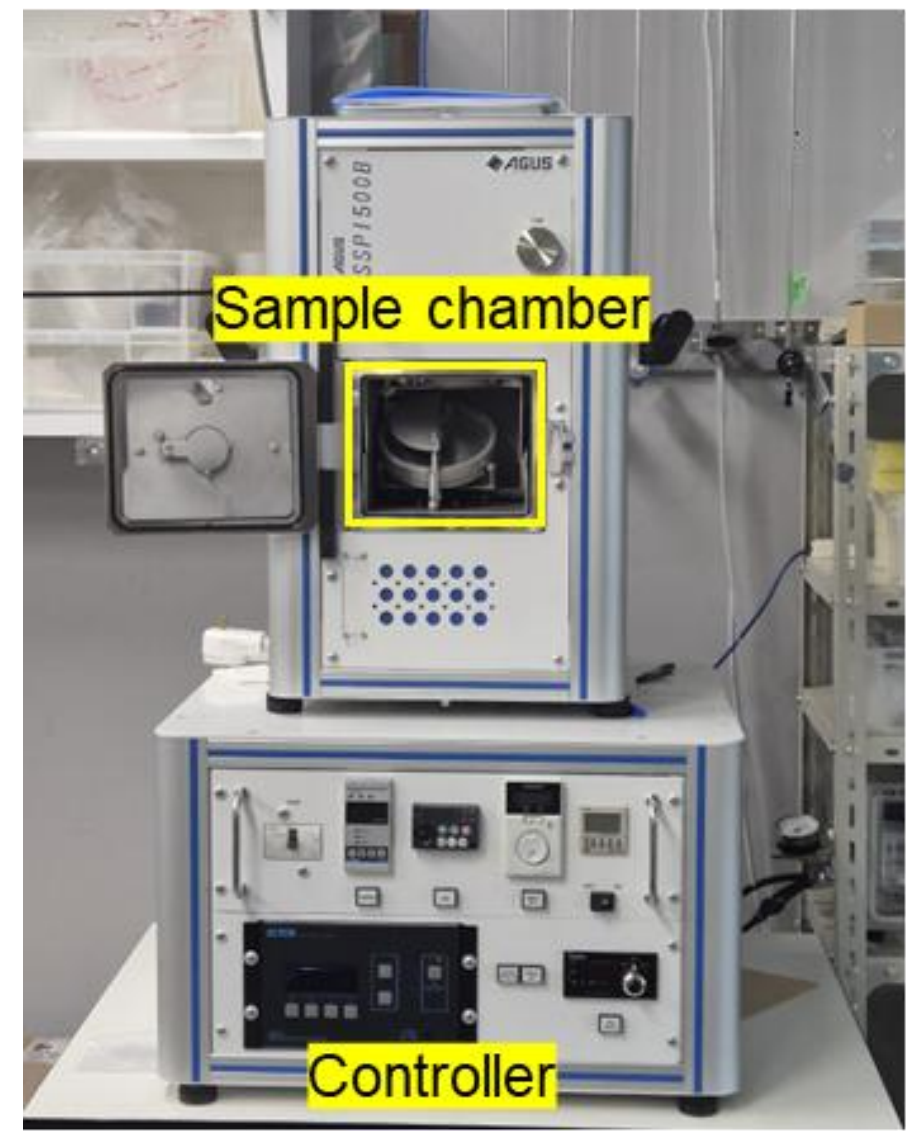

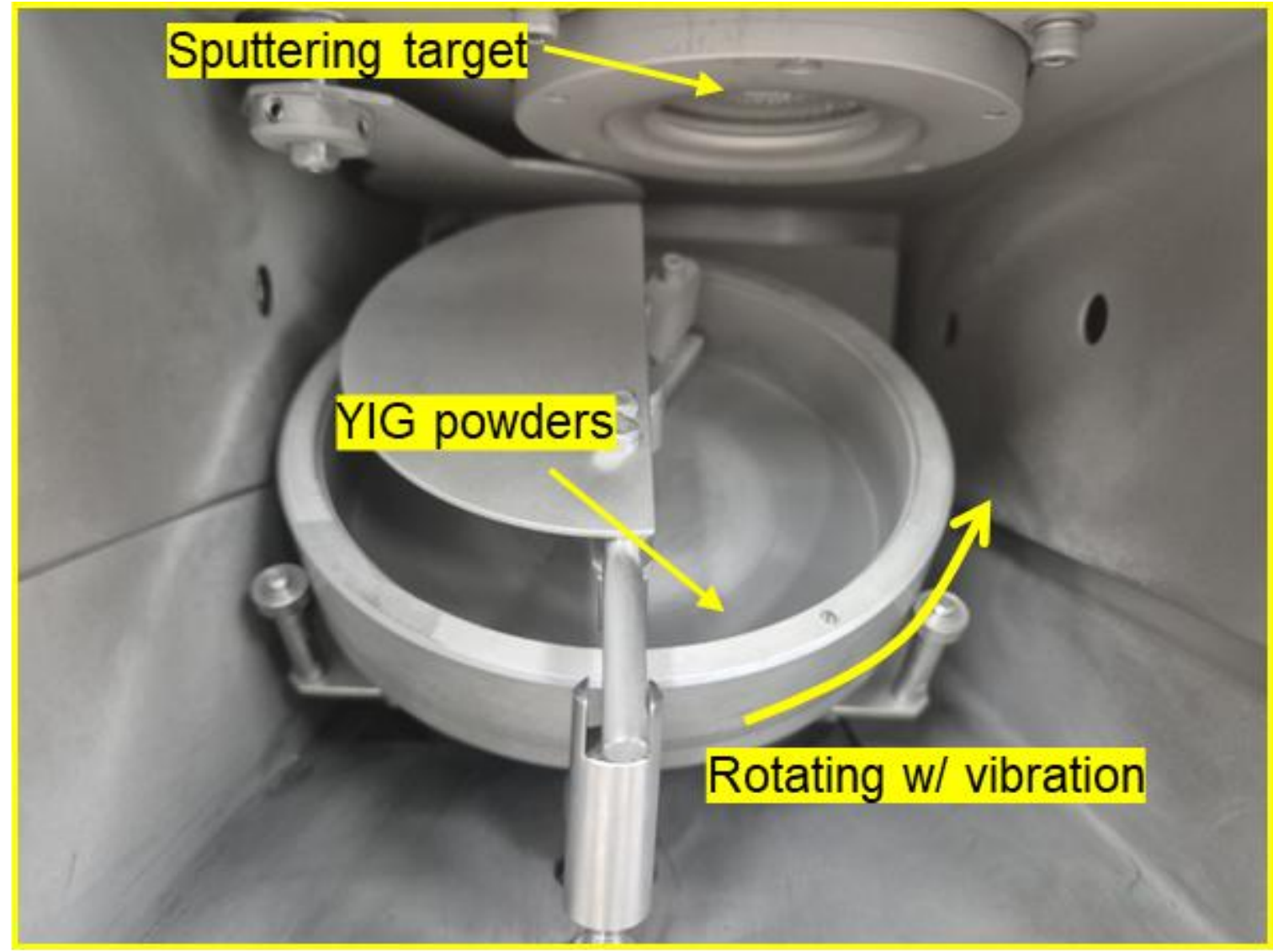

**Supplementary Fig. 1| Experimental setup for dynamic powder sputtering.**

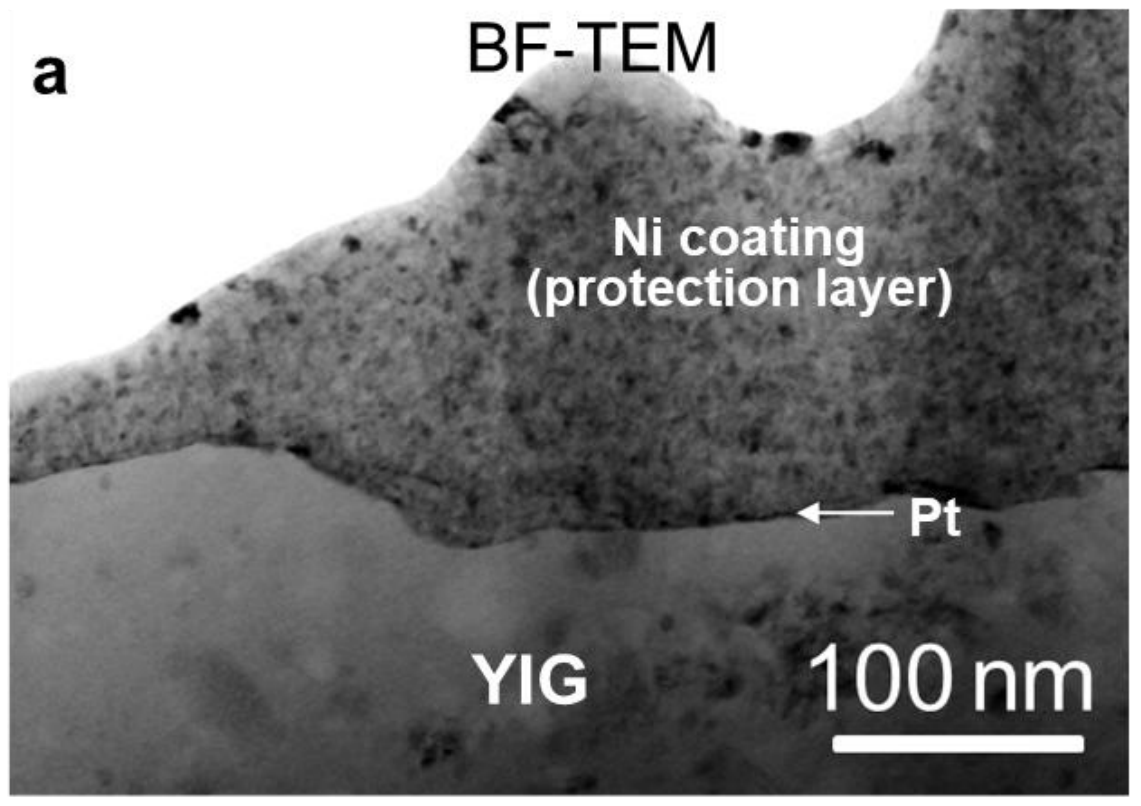

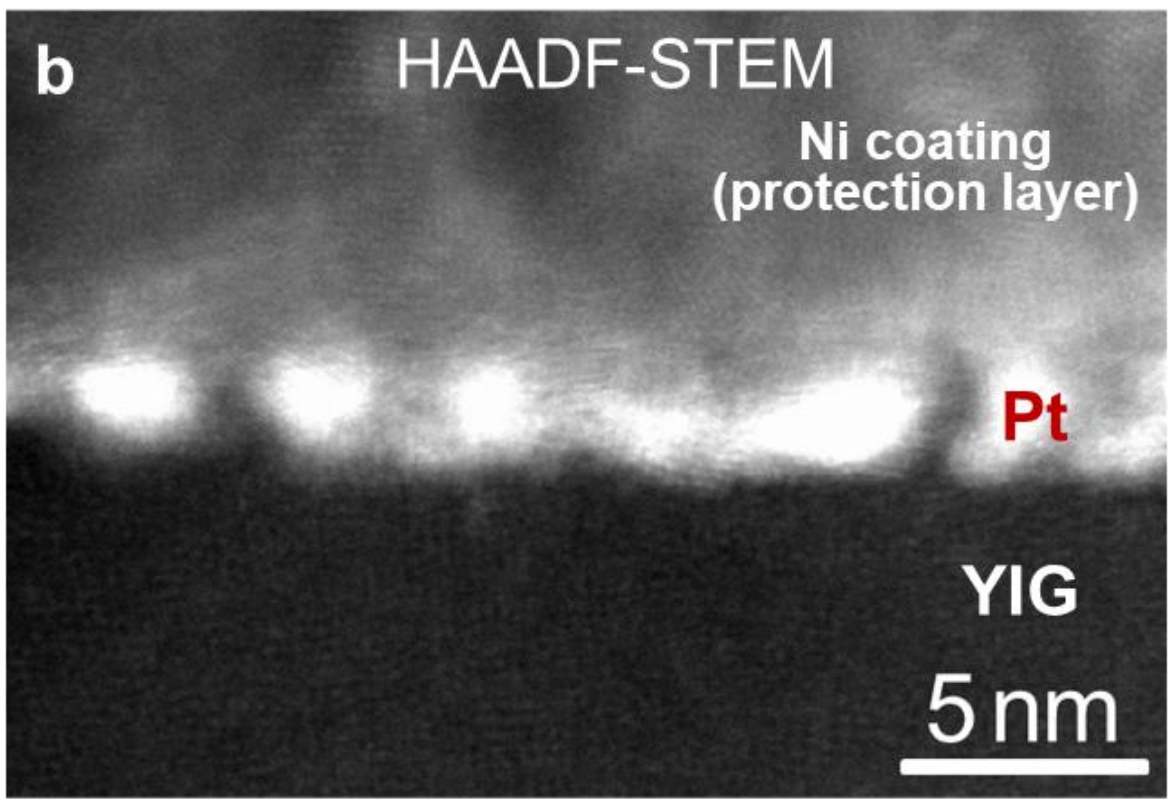

**Supplementary Fig. 2| Transmission electron microscopy (TEM) results for 2.5-nm-thick Pt-deposited YIG powders.** (a) Bright-field (BF) TEM image and (b) high angle annular dark-field scanning TEM (HAADF-STEM) image revealing island-like growth of Pt. Ni layers were deposited as a surface protective layer for TEM specimen preparation.

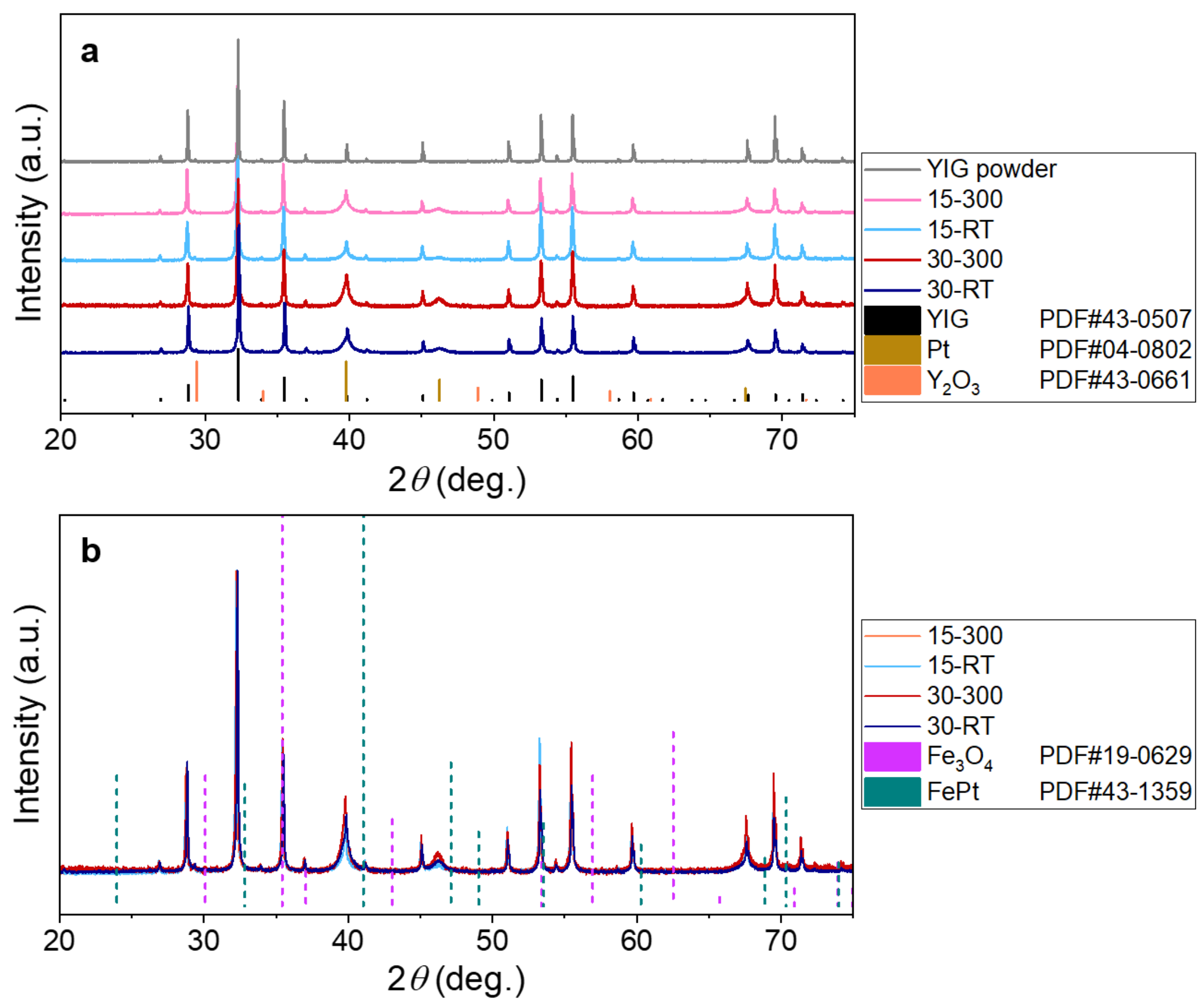

**Supplementary Fig. 3| X-ray diffraction (XRD) analysis of YIG-Pt composite samples.** (a) Comparison of XRD patterns for our samples with reference patterns of relevant constituent phases, including YIG, Pt, and $Y_2O_3$. (b) Examination of possible ferromagnetic impurity phases, such as $Fe_3O_4$ and FePt, which could contribute to the transverse thermoelectric signals.

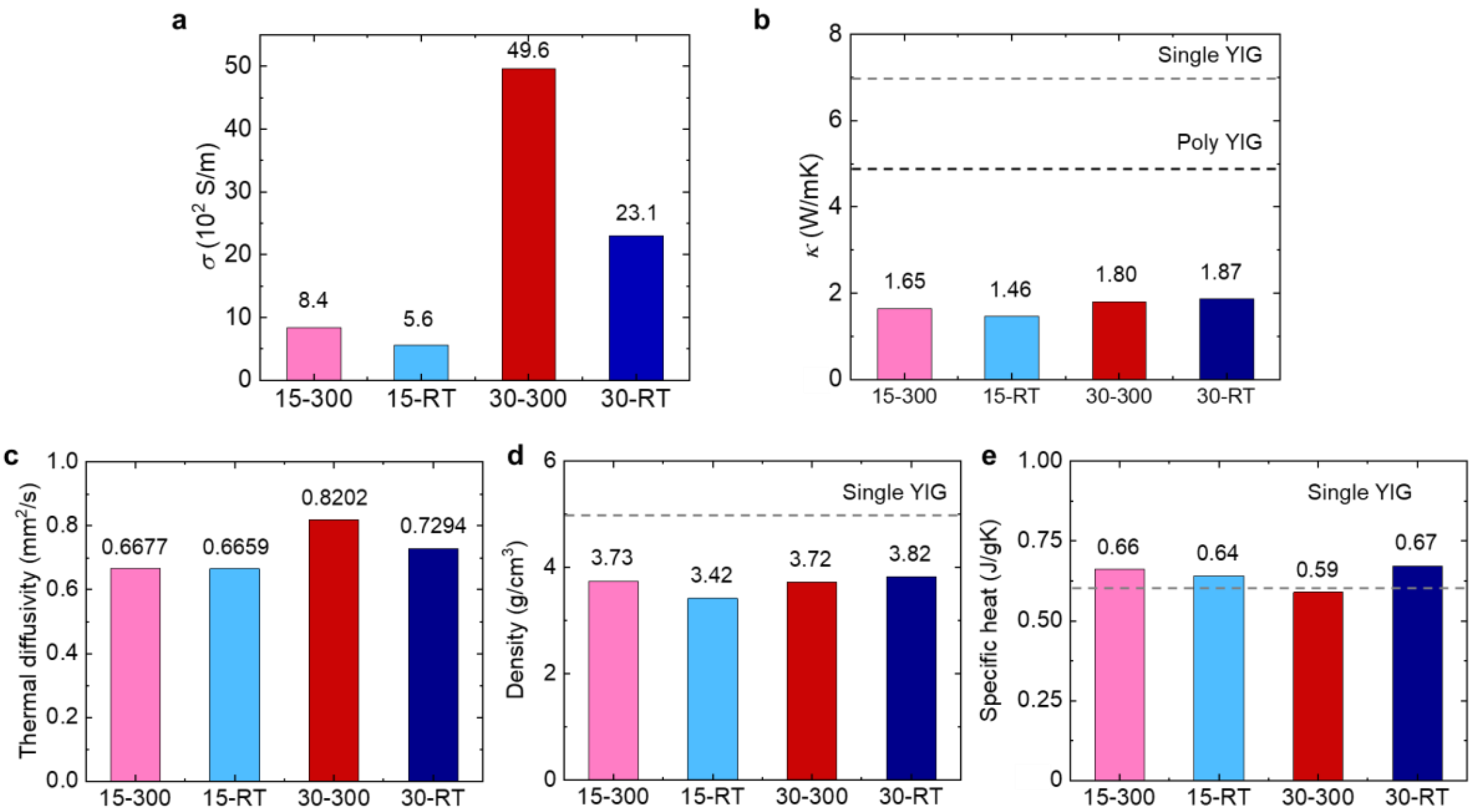

**Supplementary Fig. 4| Transport and thermal properties of YIG-Pt composite samples.** (a) Electrical conductivity ($\sigma$), (b) thermal conductivity ($\kappa$), (c) thermal diffusivity, (d) density, and (e) specific heat. The horizontal dashed lines indicate the material properties of single-crystalline or polycrystalline YIGs. The data in (a), (c)-(e) were experimentally measured, including those for the single crystal. In panel (b), $\kappa$ values for single-crystalline and polycrystalline YIGs are literature values from [1], shown for comparison.

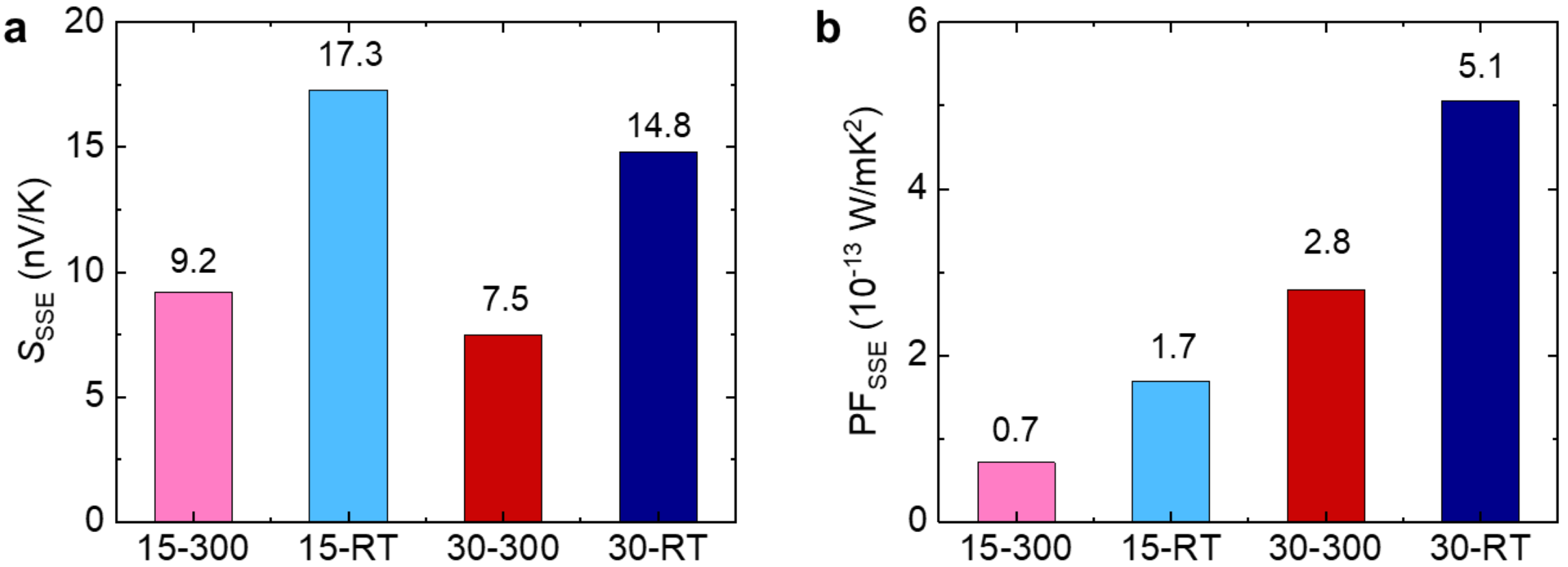

**Supplementary Fig. 5| Transverse thermoelectric properties of the samples.** (a) Spin Seebeck coefficient ($S_{SSE}$) and (b) transverse power factor ($PF_{SSE} = \sigma S_{SSE}^2$).

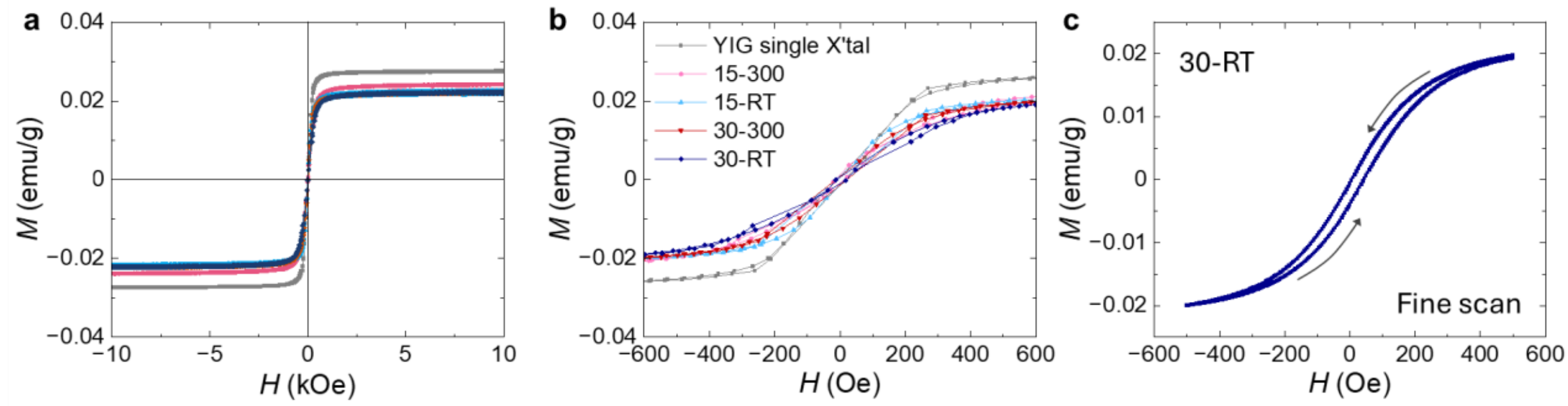

**Supplementary Fig. 6| Magnetization curves of the samples.** (a) High-field and (b) low-field magnetization data of the samples. (c) High-resolution magnetization curve of the 30-RT sample obtained with fine field steps within the low field range.

## Supplementary Table

**Supplementary Table 1. Material geometries and properties used for the maximum output power calculation.** The materials include polycrystalline $NiFe_2O_4$ (NFO), single-crystalline YIG, YIG film grown by liquid-phase epitaxy (Epi-YIG). Blanks (-) indicate properties that are not applicable.

<table>
<tr><th rowspan="2">Ref.</th><th colspan="3">Material</th><th colspan="3">Thickness</th><th rowspan="2">$\sigma$ of metallic layers (S/m)</th><th rowspan="2">$\kappa$ (W/mK)</th><th rowspan="2">$S_{SSE}$ (nV/K)</th></tr>
<tr><th>Substrate</th><th>FM</th><th>NM</th><th>Substrate (mm)</th><th>FM (mm)</th><th>NM (nm)</th></tr>
<tr><td>[5]</td><td>-</td><td>NFO polycrystal</td><td>Pt</td><td>-</td><td>0.5</td><td>10</td><td>$1.60 \times 10^6$</td><td>10.37</td><td>330</td></tr>
<tr><td>[6]</td><td>-</td><td>YIG single crystal</td><td>Pt</td><td>-</td><td>1</td><td>10</td><td>$5.30 \times 10^6$</td><td>7.9</td><td>428</td></tr>
<tr><td>[7]</td><td>GGG</td><td>Epi-YIG</td><td>$WSe_2$/Pt</td><td>0.5</td><td>0.005</td><td>5.7</td><td>$1.46 \times 10^6$</td><td>7.8</td><td>392</td></tr>
<tr><td rowspan="2">This work</td><td rowspan="2">-</td><td rowspan="2">YIG polycrystal</td><td rowspan="2">Pt</td><td rowspan="2">-</td><td colspan="2" rowspan="2">~ 1 mm</td><td>$4.96 \times 10^3$ (30-300)</td><td>1.80</td><td>7.5</td></tr>
<tr><td>$2.31 \times 10^3$ (30-RT)</td><td>1.87</td><td>14.8</td></tr>
</table>

**Supplementary References**